\documentclass[fleqn,usenatbib]{mnras}

\usepackage{graphicx}	
\usepackage{amsmath}	
\usepackage{amssymb}	

\newcommand{\Rhm}{\,$\mathrm{R}_{\textrm{H,m}}$}%
\newcommand{\msun}{\,$\mathrm{M}_\odot$}%
\def \deg{\,$^\circ$}%
\newcommand{\nb}{\,N-body}%
\newcommand{\merc}{\,\emph{Mercury}}%
\newcommand{\kep}{\,\emph{Kepler}}%
\newcommand{\mearth}{\,$M_{\oplus}$}%
\newcommand{\mjup}{\,$M_{\textrm{J}}$}%
\newcommand{\mnep}{\,$M_{\textrm{Neptune}}$}%

\title{Planet Scattering Around Binaries: Ejections, Not Collisions}

\author[Smullen, Kratter \& Shannon]{
Rachel~A.~Smullen,$^{1}$\thanks{E-mail: rsmullen@email.arizona.edu}
Kaitlin~M.~Kratter,$^{1}$
Andrew Shannon$^{2}$
\\
$^{1}$Steward Observatory, University of Arizona, Tucson, AZ 85721, USA\\
$^{2}$Institute of Astronomy, University of Cambridge, Cambridge CB3 0HA, UK\\
}

\date{Accepted XXX. Received YYY; in original form ZZZ}

\pubyear{2016}

\begin{document}
\label{firstpage}
\pagerange{\pageref{firstpage}--\pageref{lastpage}}
\maketitle

\begin{abstract}
Transiting circumbinary planets discovered by \kep\ provide unique insight into binary star and planet formation. Several features of this new found population, for example the apparent pile-up of planets near the innermost stable orbit, may distinguish between formation theories.  In this work, we determine how planet-planet scattering shapes planetary systems around binaries as compared to single stars. In particular, we look for signatures that arise due to differences in dynamical evolution in binary systems.  We carry out a parameter study of  \nb\  scattering simulations for four distinct planet populations around both binary and single stars.  While binarity has little influence on the final system multiplicity or orbital distribution, the presence of a binary dramatically effects the means by which planets are lost from the system.  Most circumbinary planets are lost due to ejections rather than planet-planet or planet-star collisions.  The most massive planet in the system tends to control the evolution.  Systems similar to the only observed multi-planet circumbinary system, Kepler-47, can arise from much more tightly packed, unstable systems.  Only extreme initial conditions introduce differences in the final planet populations. Thus, we suggest that any intrinsic differences in the populations are imprinted by formation.
\end{abstract}




\section{Introduction}

In the early part of this decade, a long-awaited discovery was made: the first transiting circumbinary planet from \kep. This planet, Kepler-16, was reported by \cite{Doyle2011}.  Since then, another eleven circumbinary planets (CBPs) have been found, including the only known circumbinary multi-planet system, Kepler-47 \citep{Orosz2012}. While the sample of planets is still small, a few unique characteristics have emerged.  \cite{Welsh2014} observe that there are no very massive, close-in planets, and the known planets tend to reside close to the stability limit of the binary.  Although these trends might arise coincidentally due to the small sample size,  if real, they hint at differences in the formation and evolution of planets around binary and single stars. In this work, we aim to tease out whether circumbinary disks might preferentially form lower mass planets near the stability boundary, or if dynamical processes sculpt the systems into what we observe.

Transiting CBPs provide important insight into planet formation and planetary dynamics because we can investigate the interplay and timeline of binary star formation and planet formation. Most simply, formation ``in-situ" around the binary is strongly favored. \cite{Armstrong2014} find that the observed CBP population is consistent with formation in a co-planar disk, unless the formation efficiency for CBPs drastically exceeds that for single stars.  This similarity aside, the formation mechanisms for CBPs may be somewhat different than those posited for planets around single stars.  Circumprimary/secondary protoplanetary disks are often truncated or less massive in close binaries, leaving less planet-forming material, while circumbinary disks can be as massive as a single-star disk \citep{Harris2012}.  In contrast, the population of the debris disks around binaries do not show flux deficits, as might be expected given the reduced mass in the parent population \citep{Rodriguez2015}. \cite{Martin2013} propose that CBP formation might happen more efficiently in dead zones (quiescent regions in the disk mid-plane), which could produce gas giants easily.  On the other hand, the binary can excite substantial eccentricity in the protoplanetary disk, inhibiting planet formation near the binary. The eccentric disk gives rise to eccentric planetesimals which suffer high velocity collisions that lead to erosion instead of growth, pushing planet formation to larger radii (\cite{Marzari2013}, \cite{Silsbee2015}) . If  a planet instead forms in the outer disk and migrates inward due to tidal interactions with the disk, one might still expect planets to exist close to the stars, which \cite{Bromley2015} posit to be the likely scenario. \cite{Pierens2013} find that a planet forming in the outer disk can migrate toward the stability limit but will probably be pumped to moderate eccentricity along the way.  Additionally, \cite{Pierens2008} find that massive planets, if they exist around binaries, are probably found at larger radii because tidal torques from the binary cause outward migration. 

We must also consider planet formation in the presence of binary evolution. There is a lack of observed planets around short period binaries  (periods less than about 7 days; \cite{Armstrong2014}, \cite{Martin2014}).  Models proposed in \cite{Mazeh1979} and \cite{Fabrycky2007} suggest that these binaries form on wider orbits and then migrate due to tidal circularization stemming from Kozai oscillations induced by a tertiary companion.  \cite{Martin2015d} suggest that this is prohibitive for CBP existence around a tight binary, while \cite{Munoz2015} and \cite{Hamers2016} posit that CBPs may just become very misaligned.  Both of these scenarios would provide a dearth of transiting CBPs around close binaries.

Formation alone, however, does not explain the present-day orbits in planetary systems around single stars.  Previous works, such as \cite{Chambers1996}, \cite{Faber2007}, \cite{Juric2008}, \cite{Chatterjee2008}, \cite{Smith2009}, \cite{Raymond2010}, \cite{Lissauer2011},  and \cite{Pu2015} have looked at the impact of planet scattering on planet populations around single stars.  \cite{Mustill2014} and \cite{Veras2014} have extended this to understand dynamical evolution over the full stellar lifetime.  Specifically, \cite{Juric2008}, \cite{Chatterjee2008}, and  \cite{Pu2015} have found that the observed exoplanet sample is consistent with significant sculpting by dynamical evolution. This naturally raises the question of how scattering is modified around binaries. 

The addition of a second massive body substantially changes stability very close to the binary.  \cite{Holman1999} have shown empirically that orbits within about two times the binary semi-major axis are unstable on very short timescales, suggesting that neither planets nor the natal disk should exist in this region.  However, it is unclear how significantly planets on wider orbits will be impacted except at special locations such as mean motion resonances with the binary. One possible avenue of further evolution is the modest eccentricity excitation at semi-major axis 2-10 times the binary semi-major axis, which may fundamentally change the course of planet-planet scattering and thereby change the resultant population. In this work we aim to understand the impact of the binary on planet populations sculpted by planet-planet scattering. By isolating the role of the binary in any differential evolution due to scattering, we can determine which differences are imprinted by formation.

To address the interplay between the formation and dynamical evolution of circumbinary planets, we perform N-body integrations of planets around single and binary stars. We study the binary's impact on a wide range of different planet populations, investigate the changes in orbital properties as a result of dynamical processes, and compare the resultant populations around single and binary stars. We first review previous work in Section~\ref{stab}. In Section~\ref{methods}, we discuss the methods used to carry out our study and explain our choice of systems and planet populations. Section~\ref{results} details the differences we see between the various planet populations and between planetary systems around single and binary stars. Section~\ref{disc} discusses the physical intuition for the reduction of collisions, the role of giant planets in system evolution, and the observable properties of our systems.

\section{Planetary Stability}\label{stab}

While any system of three or more bodies may be chaotic, there are several limiting cases where orbits are well behaved.  The simplest case is that of two planets around a single star that are Hill stable, which means that they cannot suffer close encounters. \cite{Gladman1993} explored Hill stability for low mass, low eccentricity, co-planar bodies and found that systems of two planets are Hill stable for orbital separations greater than $\Delta>2.4((m_1+m_2)/M_{*})^{1/3}$ where $m_1$ and $m_2$ are the masses of the planets and $M_{*}$ is the mass of the central star; here, the orbital radius of the inner planet is taken to be 1. 

Multi-planet stability is often referenced to the two-planet Hill stability limit by measuring planet spacing in terms of  a mutual Hill radius:
\begin{equation}\label{eq:mhr}
R_{\textrm{ H,m}}=\left( \frac{m_{1}+m_{2}}{3M_{*}} \right)^{\frac{1}{3}}\frac{a_1+a_2}{2}
\end{equation}
where $a_1$ and $a_2$ are the semi-major axes of the planets.  We define the dimensionless spacing of planets in terms of mutual Hill radii as: 
\begin{equation}\label{eq:beta}
\beta=\frac{a_2-a_1}{R_{\textrm{ H,m}}}
\end{equation}
 Note that in  some regimes, $\beta$ may not provide the best metric for planetary stability (see \cite{Morrison2016}).  Previous works such as \cite{Chambers1996}, \cite{Faber2007}, \cite{Smith2009},  \cite{Shikita2010}, \cite{Lissauer2011}, and \cite{Pu2015} have studied the impact of $\beta$ on the dynamical ``lifetime,'' meaning the timescale for planets to enter crossing orbits, for systems of three or more equal mass planets around a single star.  They find that the lifetime of a system decreases with increasing planet mass, planet eccentricity ($e$), and system multiplicity, and increases  with the initial spacing measured by $\beta$.  \cite{Chambers1996} suggested that Gyr stability requires $\beta>$10 for $>3$ planet systems and \cite{Smith2009} found that a spacing of  $\beta>$8 is required for Myr stability in systems with five or more equal mass planets. \cite{Kratter2014} investigated two-planet circumbinary systems and found that they are long-term stable ($10^8$ binary orbits)  with  $\beta>$7. Because we are interested in CBPs of higher multiplicity and non-constant mass, we therefore might expect our planet distributions to require larger spacing than this in order to be stable for tens to hundreds of millions of binary orbits.

\section{Methods}\label{methods}

\subsection{Integrator}\label{int}

 Our integrations are carried out with a Gauss-Radau variable timestep integrator in a modified version of the \nb\ orbital integration package \merc\ from \cite{Chambers1997}. The standard variable time step orbit integrators included in the code such as Bulirsch-Stoer and Gauss-Radau are agnostic about the number of massive bodies or hierarchy of the system, and therefore are well suited to planet-binary integrations in general \citep{Youdin2012,Kratter2014,Sutherland2016}. While binary symplectic integrators exist \citep{Chambers2002,Beust2003}, these still require switching to a B-S style integration to resolve close encounters.  If encounters are common, the integrator will be forced to use B-S integration schemes for a significant fraction of the integration.

Because of \merc 's origins as a planetary system integration package, most calculations are carried out in heliocentric coordinates. While this poses no challenge for the main \nb\  integration for some integrators, any part of the code which relies on assumptions of Keplerian orbits about a central body requires modification, such as the close encounter checks. The changes to \merc\ described herein remove this assumption when the user sets a flag for a central binary in one of the input files. This new version of the code is available for download online.\footnote{\texttt{https://github.com/rsmullen/mercury6\_binary}} We briefly describe the main modifications below. All of our integrations were carried out with the \cite{Everhart1985} Radau integrator, although the modifications should work with other adaptive time step methods.

\begin{itemize}

\item{For circumbinary systems, we treat close encounters between any two bodies in the same way, in contrast to the standard \merc\ practice of treating encounters with the central star separately. For any pair of bodies, the code searches for close encounters based on the current Cartesian state vectors. For the Radau integrator, interactions flagged as close encounters do not effect the overall time stepping in the code. This is in contrast to hybrid symplectic integrators that use a close encounter flag to choose interactions to further resolve. For the Radau integrator, the variable time step ensures that interactions down to the close encounter radius of a particle are well resolved. For planet encounters, we use 1 $R_{\rm Hill}$ as the close encounter radius, following previous work \citep{Juric2008}. For stars,  we use the empirical stellar mass-radius relationship from \cite{Demircan1991} to determine the radii as a function of mass. We set the close encounter radius to three stellar radii for our fiducial runs. Note that the central body radius and second star's radius are set in the subroutine ``\texttt{mfo\_user\_centralradius},'' which can be easily modified to incorporate any prescription.}

\item{Collisions, like close encounters, are also calculated for every pair of bodies based on Cartesian state vectors. They are calculated based on extrapolation from the close encounter radius over a time step. We use \merc 's third order interpolation scheme for all bodies, which ignores the gravitational contributions of all other bodies during the encounter. For planet-planet encounters, the choice of close encounter radius between 1/4 and 1 $R_{\rm Hill}$ does not change the number of planet-planet collisions, so we have chosen the default of 1 $R_{\rm Hill}$ to avoid the computational cost of very small time steps. For star-planet encounters, where ignoring the gravitational accelerations from one of the stars is most severe, the time step is always sufficiently short compared to the orbital period that the star moves of order 1 stellar radius during the extrapolated encounter. The time step is guaranteed to be small, ~1/1000th of an orbital period, because the small close encounter radius for stars forces very high time resolution during close approaches. To achieve even better accuracy for stellar collisions, one can set the close encounter radius equal to the collision radius to force the integrator to resolve all collisions explicitly. We find that the number of stellar collisions for an equal mass binary is exactly the same for close encounter radii from 1 to 3 stellar radii; however, because of the inherent chaos of the systems and the different time resolutions, the time of collision, and the binary component that suffered a collision, may change.}

\item{The standard \merc\ routines for calculating Jacobi coordinates and planetary Hill radii must also be modified to account for both re-ordering of planets and the offset between the system center of mass and the central body. We include a new Jacobi coordinate routine that employs a bubble sort algorithm to re-order bodies by distance before performing a coordinate transform from heliocentric coordinates. Hill radii for close encounters are calculated using the distance from the system center of mass, rather than semi-major axis, and incorporate the enclosed mass instead of the mass from a single central body.}

\item{Finally, we apply the bug fix reported in \cite{DeSouzaTorres2008} that fixes a status initialization problem, although this is not used in the current study.}

\end{itemize}

Using this modified version of \merc, we obtain an average energy conservation of about $10^{-5}$ over our 10 Myr ($3\times 10^8$ binary orbits) integration time, with ranges from $10^{-7}$ to $10^{-5}$.  Our angular momentum conservation is of order $10^{-5}$.  \cite{Juric2008}, who also used a custom version of \merc\, found an energy error of up to $10^{-4}$ using their hybrid symplectic/Bulirsch-Stoer scheme.

We have also performed a code comparison with another publicly available integrator, \textsc{Rebound}, \citep{Rein2012}. We employ their 15th order integration scheme which is similar to Radau, but conserves energy significantly better \citep{Rein2014}. The trade off is of course a dramatic (order of magnitude) increase in run time, which made it unfeasible for use in this parameter study. While \textsc{Rebound} automatically treats encounters between any pair of bodies equivalently and operates in barycentric coordinates, one must modify collision routines and ensure that the system does not drift out of the box by frequently resetting the system back to the center of mass. For integrations of identical initial conditions drawn from our fiducial sample we achieved very good agreement between the two integrators (0.1\% difference for planet-planet and planet-star collisions and 1.7\% difference in ejections and remaining planets when the systems are run to $10^5$ years). Note that, due to the highly chaotic nature of the orbits, numerical errors introduced by the different integration algorithm are expected to produce small changes in orbit outcomes for a given planet. Additionally, the ejection  algorithms, in particular, are different (\merc\ ejects from a sphere, while \textsc{Rebound} ejects from a cube), so we expect small differences in the outcomes from these effects, as well. 

\subsection{Planetary Initial Conditions}\label{ics}

There are two major influences on the long term evolution of CBPs: the properties of the binary system and the structure of the planetary system.  Because the primordial conditions of circumbinary systems are uncertain, we consider a range of planet populations and binary configurations to investigate the dependence upon initial conditions.  We do not use the observed planetary statistics for our populations because observed systems may already be sculpted by scattering.   

Our systems are initialized with ten planets that have been randomly drawn from the planet populations described below. Our fiducial binary is equal mass and circular with components of 0.5\msun\ separated by 0.1~AU, which gives a 10 day period. We make no assumptions about stability other than re-sampling any planet that falls within the \cite{Holman1999} circumbinary stability limit of $a_{\textrm {pl}}< (2.278+ 3.824e_{\textrm b}-1.71e^2_{\textrm b})a_{\textrm b}$ (equation 5 in their text) where $a_{\textrm {pl}}$ is the planet semi-major axis computed from the system barycenter and $a_{\textrm b}$ and $e_{\textrm b}$ are the semi-major axis and eccentricity of the binary.  This resampling forbids initial conditions interior to the probable disk truncation edge. Planets are unlikely to have formed or even migrated into this region and may contribute to overall system destabilization, thus polluting our statistics.  We apply the same initial semi-major axis cutoff for planets around a single star as we do for the binary to ensure consistency between our populations, although no such restrictions exist around single stars.  Thus, very short period planets are initially forbidden around single stars but can be scattered inward during the simulation.

Each distribution is integrated with both our fiducial binary and a single star with a mass of 1\msun\, making the effective central mass in both cases the same. For a subset of planet populations, we vary the binary eccentricity and/or mass ratio (we define mass ratio as $\mu=M_2/M_1$); these variations are listed in Table~\ref{tab:pcts}.  Although we only study one binary orbital period herein, our results are mostly scalable to wider periods, as we discuss in Section \ref{close}.

For all populations, we assume the eccentricity and inclination ($i$) distributions of \cite{Juric2008}, who used a Rayleigh distribution with scale parameter $e=0.1$ and $i=5.73$\deg.  Inclinations are somewhat uncertain, so we remain with the low inclination distribution to represent a mostly flat disk formation scenario, such as suggested in \cite{Fang2012}.  The eccentricity we use is roughly consistent with the observations of \cite{VanEylen2015a}, who find that  observed planets follow a Rayleigh distribution with scale parameter $e=0.05$. 

Our mass and semi-major axis distributions are described below.  All orbital elements henceforth are described with respect to the system barycenter.

\begin{figure}
\centering
\includegraphics[scale=.25]{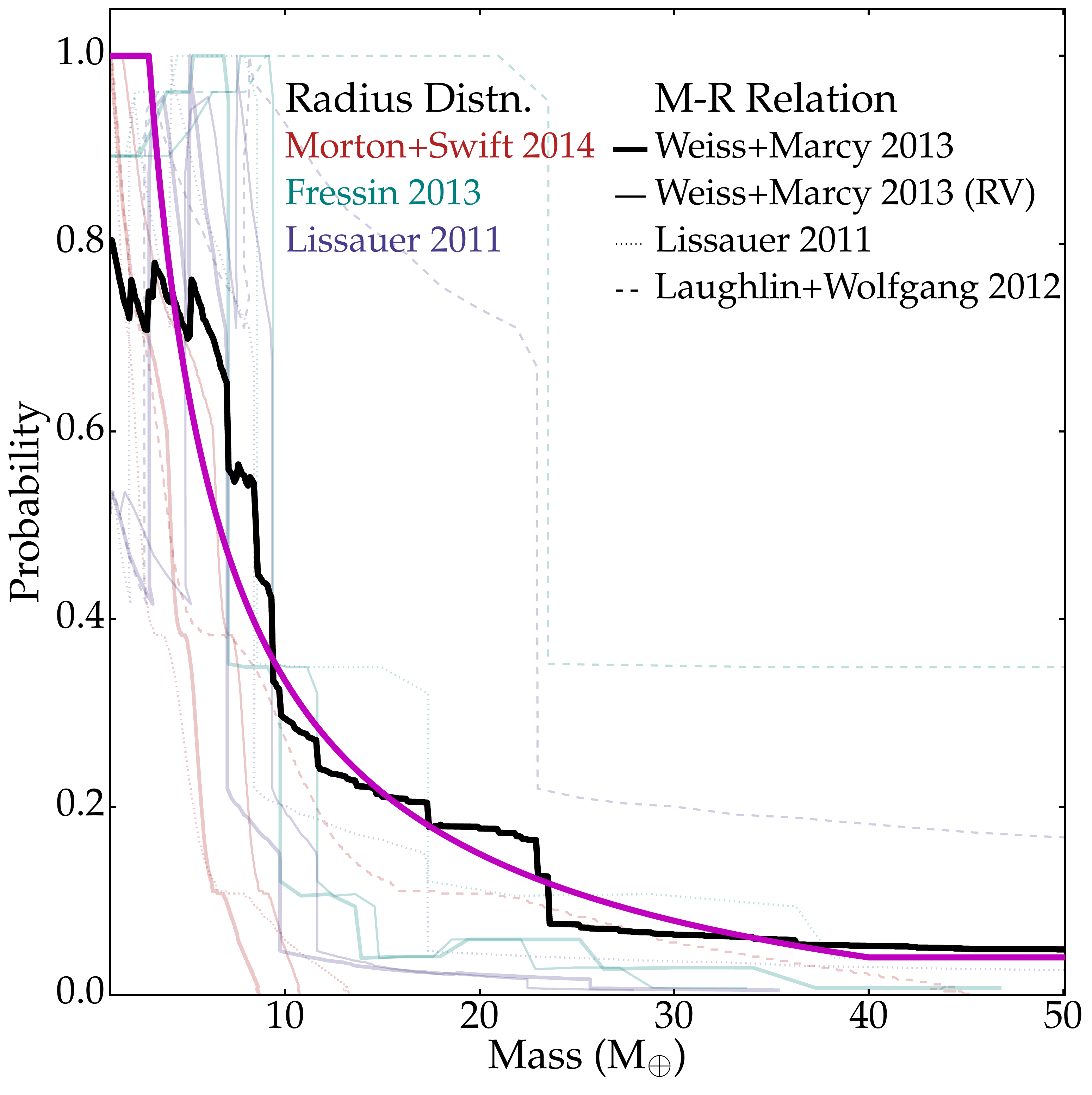}
\caption{ The derived probability distribution for planet masses in the LM planet sample. Each line indicates a probability distribution of planet mass in Earth masses.  Each of the three colors (red, teal, and blue) represents a different completeness-corrected radius distribution derived from the \kep\ planet sample. The different line styles denote the mass-radius relation applied to each observed radius distribution. The thick black line is the average of all of the red, blue, and teal lines. The magenta line shows our analytic expression for the probability distribution using an exponential function with flat probabilities at $M<3$ \mearth\ and $M>40$ \mearth\ as shown in equation 3.   We use this mass distribution to produce an observationally motivated planet sample.
\label{lmmass}}
\end{figure}

\begin{description}
\item[{\textbf{JT08}}] This distribution serves as our reference sample and is taken directly from the \cite{Juric2008} ``c10s10'' ensemble to compare CBPs to previous simulations around single stars. Planet masses are drawn from a log uniform sample ranging from 0.1 to 10 \mjup\ and semi-major axes are drawn from a log uniform sample between 0.1 and 100~AU.  

\item[{\textbf{MMHR}}] The MMHR (matched mutual Hill radius) distribution matches the initial planet-planet spacing of JT08, as measured by the mutual Hill radius spacing, but with lower mass planets and smaller semi-major axes.  The planets are drawn from a log uniform distribution spanning 1 to 160 \mearth\ in mass and log uniform from 0.1 to 1.7~AU in semi-major axis. We choose this population to highlight the impact of binary-planet perturbations. The dynamical spacing of planets (a measure of the strength of inter-planet perturbations) is the same as in JT08, but binary perturbations will be stronger because of the compact nature of the population.

\item[{\textbf{Mordasini}}] This planet population is modeled after the population synthesis models of \cite{Mordasini2009a,Mordasini2009b}.  The semi-major axes span a range from 0.1 to 15~AU with a peak at 3~AU.  The masses span 1--$10^4$ \mearth, with a dominant peak at low mass and small peaks around 1 \mnep\ and 1 \mjup.  The Mordasini data from which our distributions are taken do have a correlation in mass-semi-major axis space; however, our randomly drawn planets do not take this two-dimensional density into account. 

\item[{\textbf {LM}}] In order to capture the properties of observed exoplanets around single stars, even though the present-day distribution may not be primordial, we create the LM (low mass) planet population.  We apply the empirical planetary mass-radius relations from \cite{Weiss2014}, \cite{Lissauer2011}, and \cite{Wolfgang2012} to observed radii distributions from \cite{Morton2014}, \cite{Fressin2013}, and \cite{Lissauer2011} to create an average mass distribution for exoplanets.  We then match an analytic expression for the probability distribution function using an exponential with flat probabilities at $m<3$ \mearth\ and $m>40$ \mearth, as shown in equation~\ref{eq:lm}.

\begin{equation}\label{eq:lm}
P(m)=\begin{cases}
1;& m<3\textrm{ \mearth}\\
\left(2.758\times m^{-0.745}-0.133\right)/1.083;& 3\textrm{ \mearth}<m<40 \textrm{ \mearth}\\
\left(2.758\times 40^{-0.745}-0.133\right)=0.043;& m>40 \textrm{ \mearth}
\end{cases}
\end{equation}

 Figure~\ref{lmmass} shows the average mass distribution in the thick black line and the analytic fit used for the LM distribution in the thick magenta line. We draw masses from 1--$10^4$ \mearth;  most planets have masses less than 20 \mearth.  The semi-major axes are drawn from a gamma distribution with mean 4.5~AU and range approximately 0.1--15~AU.  
\end{description}

\subsection{Integration Parameters}

We integrate 100 different realizations for each system architecture for 10 Myr. Each system begins with 10 planets. A planet is considered ejected if it travels more than 1000~AU from the primary star.\footnote{We do not account for the offset of the primary from the system center of mass, as this distance is negligible in comparison to the ejection radius.} Note that, for the widely spaced planet populations, there is a small subset of high eccentricity, high semi-major axis planets that are removed from the system while still bound.  However, these planets are a minority and would likely not contribute greatly to the further dynamical evolution of the system. Collisions between all objects are allowed in a mass and momentum conserving form (we do not allow collisional erosion or tidal dissipation). 

Planetary radii are calculated using mass and an assumed density of $\rho=1$g/cm$^3$.  This assumed density best describes a normal giant planet, such as Jupiter. For the range of densities of known \kep\ systems, we underestimate the radius by at most a factor of two for the least dense planet and overestimate by a factor of three for the most dense planet.  Our assumption of a constant density should have negligible impact on planetary collisions; in fact, because most of our planets are smaller than Jupiter, we should more frequently overestimate the radii and therefore overestimate collisions.  

It is important to note that this problem is scale free aside from collisions, which of course set an absolute radius. Otherwise, binary and planetary orbits can be scaled up or down, with the timescale adjusted accordingly.  Because we find that collisions of any kind are relatively rare for most distributions, the trends presented here should be applicable to wider circumbinary systems. In these systems one would expect collisions to occur even less frequently due to increased distances between objects.

\section{Results}\label{results}

 The primary difference we observe between planetary systems around single and binary stars is the loss mechanism of unstable planets.  Circumbinary planets are lost almost exclusively by ejections, whereas single star planetary systems undergo a substantial number of planet-planet and planet-star collisions.  The evolution at 10~Myr has reached a near steady state; most systems have non-crossing planetary orbits that have changed little over the last few Myr. Figure~\ref{nvt} shows the average number of planets remaining in the system as a function of time.  Figure~\ref{pies} shows the outcomes for planets in the four populations. Despite the differences in outcome, single stars and binaries asymptote towards similar orbital distributions, except for the most compact, packed initial populations.  

Though the final distributions of orbital elements for each planet population are relatively invariant with central object, each planet population retains a ``memory'' of its original state, which can be seen in Figure~\ref{bvs};  the shapes of the final distributions vary significantly between different initial populations.  

Scattering does not appear to account for the pileup of observed planets near the stability limit, nor do binaries preferentially lose massive planets close to the binary. Scattering thus does not reproduce these noticeable features of the observed CBP population. However, the fate of planets that begin or are scattered close to the binary star (within $\sim10a_b$) is different from those that never enter this region, as we explore in Section~\ref{close}. We discuss in Section~\ref{massive} that the presence or absence of a giant planet has greater impact on the dynamical evolution of a system than the central object. 

\subsection{Differences between single and binary planet populations}

\begin{figure}
\centering
\includegraphics[scale=0.35]{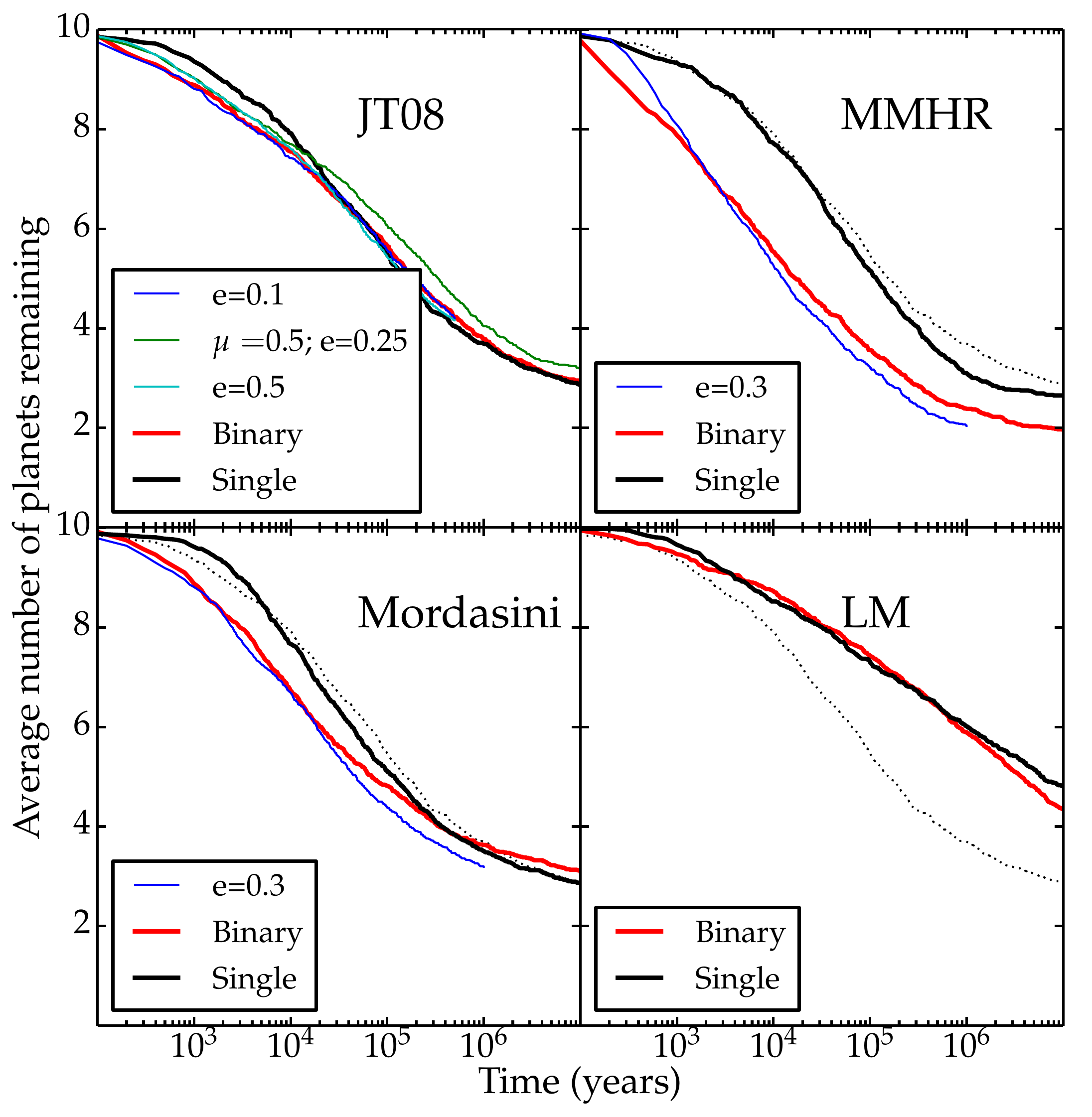}
\caption{ The average number of planets remaining per system as a function of time for each planet distribution.  Red lines show the fiducial equal-mass, circular binary planetary systems and black lines show the single star planetary systems.  Other colors show variations of the central binary (in $e$ or $\mu$).  The dotted black line shows the JT08 single star case as a reference.  In every case but MMHR, the single and binary systems lose very similar numbers of planets.  However, the rate of loss can vary, with the binary systems tending to lose planets faster at early times. In contrast to the other distributions, the LM planet population is still dynamically evolving at 10 Myr because low mass planet-planet encounters rarely lead to ejections. 
\label{nvt}}
\end{figure}

The four planet distributions around the fiducial binary show a factor of $\sim 20$--30 reduction in planet-star collisions.  Planet-planet collisions are reduced by 1--2 times, and ejections are enhanced by factors of 1.5--2.5. We provide a physical explanation for these differences in Section~\ref{explanation}. The average number of planets remaining in a system as a function of time is generally similar between the single star and the binary case. Systems reach 10~Myr with 2--4 planets remaining, on average. Figure~\ref{nvt} shows the time evolution of the number of planets in the system. 

\begin{figure*}
\centering
\includegraphics[scale=.45]{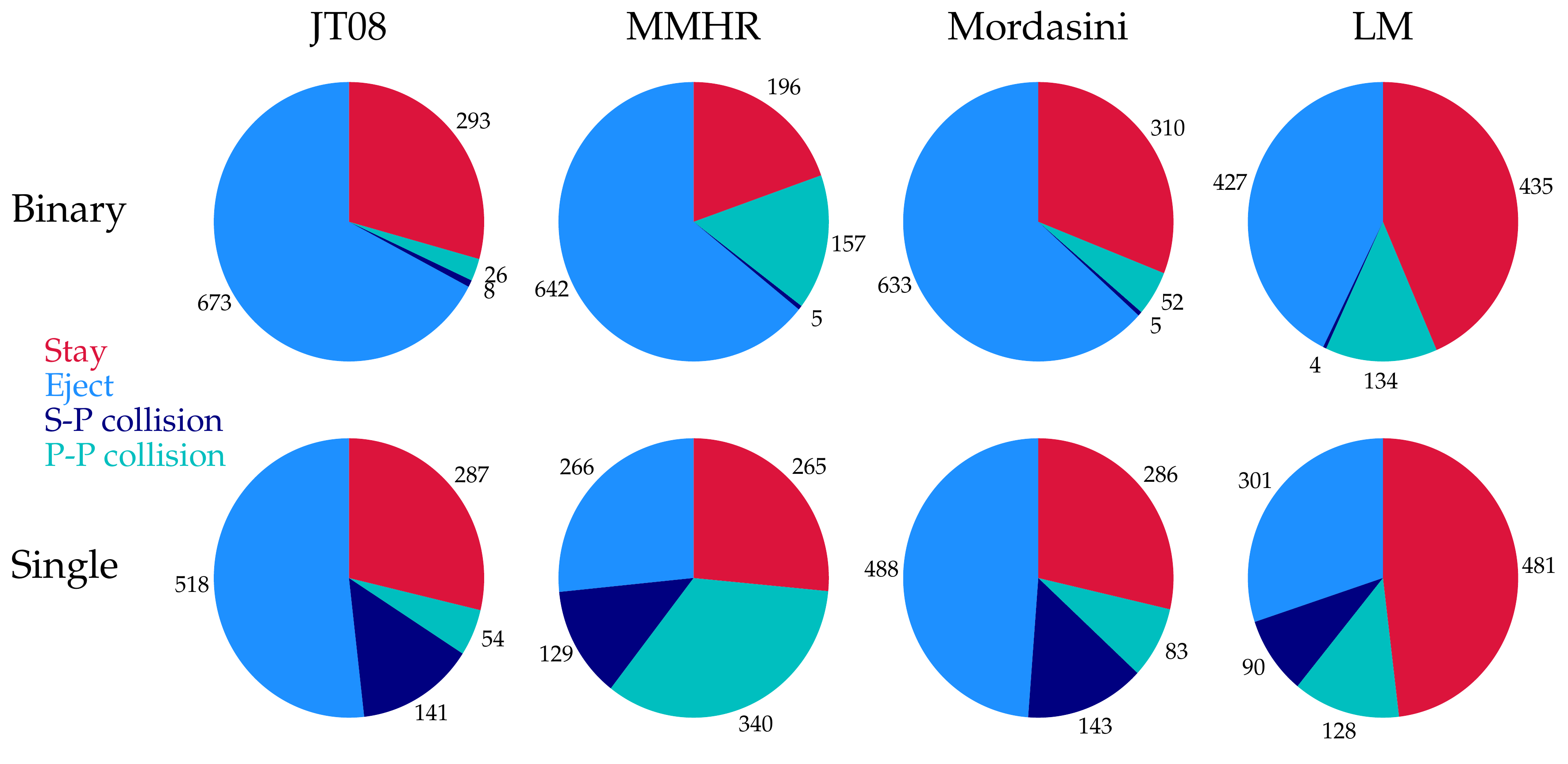}
\caption{ Pie charts showing the fate of planets in each distribution at 10~Myr for both binary (top row) and single (bottom row) stars.  The columns correspond to the JT08, MMHR, Mordasini, and LM planet populations from left to right.  Going counterclockwise, red shows the fraction of planet remaining in each system.  Blue, navy, and cyan indicate the fraction of planets lost by ejections, stellar collisions and planet-planet collisions, respectively.  Although the number of remaining planets is roughly constant between single and binary systems, the loss mechanism is very different. The planets around binaries suffer far fewer collisions in exchange for far more ejections.
\label{pies}}
\end{figure*}

\begin{description}
\item[{\textbf{JT08}}] The JT08 set of initial conditions is the only one in which we have a direct comparison to previous work. We present the results of the \cite{Juric2008} ``c10s10'' integration alongside our own in Table~\ref{tab:pcts} and see that the single star integrations are consistent. 

Over our 1000 planet sample for each of the binary and single star tests, we found a nearly equal number of remaining planets, an enhancement of 1.3 in ejection rates for the binary case, a reduction of 17.6 times in the number of planet-star collisions, and a factor of 2.1 reduction in planet-planet collisions.  The loss mechanism is shown graphically in the first column of Figure~\ref{pies}, which depicts the fractional distribution of planet outcomes at 10~Myr. The average number of planets remaining at the end of the integration times is also shown in Table~\ref{tab:pcts}, which lists both the fiducial binary ($e=0; \mu=1$) and variations on binary eccentricity  and mass ratio.  Because we don't run all variations on the binary to 10~Myr, we include data from the fiducial binary at the shorter times for comparison.   We find that the differences between planet populations around different binaries are small. 

\item[{\textbf{MMHR}}] In the binary population, we found a reduction of 1.4 times in the number of planets remaining, an increase of 2.4 times the ejection rate, and reductions of 25.8 and 2.2 in planet-star and planet-planet collisions relative to the single star case.  The large increase in ejections in the binary case can be attributed to the compactness of the population (the median MMHR semi-major ax is $\sim$0.6~AU or $6a_b$, as opposed to  $49 a_b$, $20a_b$ and $22a_b$ for JT08, Mordasini, and LM, respectively).  The fractional loss rates for this set of initial conditions can be seen in the second column of Figure~\ref{pies}, and the average number of planets remaining in the system can be seen in Table~\ref{tab:pcts}.

\item[{\textbf{Mordasini}}] Despite having wildly different planet mass and semi-major axis distributions, the Mordasini population behaves most similarly to the JT08 population. On average, each Mordasini system will have one giant planet ($M>M_J$), which leads to the similarities in evolution (see Section~\ref{massive}).   Comparing the binary and single star planet populations, there is no significant change in the number of planets remaining, a factor of 1.3 more ejections for CBPs, and reductions of 28.6 and 1.6 in planet-star and planet-planet collisions, respectively. The outcomes of planets in this integration are shown in the third column of Figure~\ref{pies} and in Table~\ref{tab:pcts}. 

\item[{\textbf{LM}}] This is our least active sample due to the wide initial spacing in mutual Hill radii and low average planet mass.  The results for this planet population should be interpreted with caution because about 50\% of systems are still undergoing significant dynamical evolution at 10Myr. We show the comparison of binary and single outcomes in the fourth column of  Figure~\ref{pies} and in  Table~\ref{tab:pcts}.  

\end{description}

\begin{figure*}
\centering
\includegraphics[scale=.3]{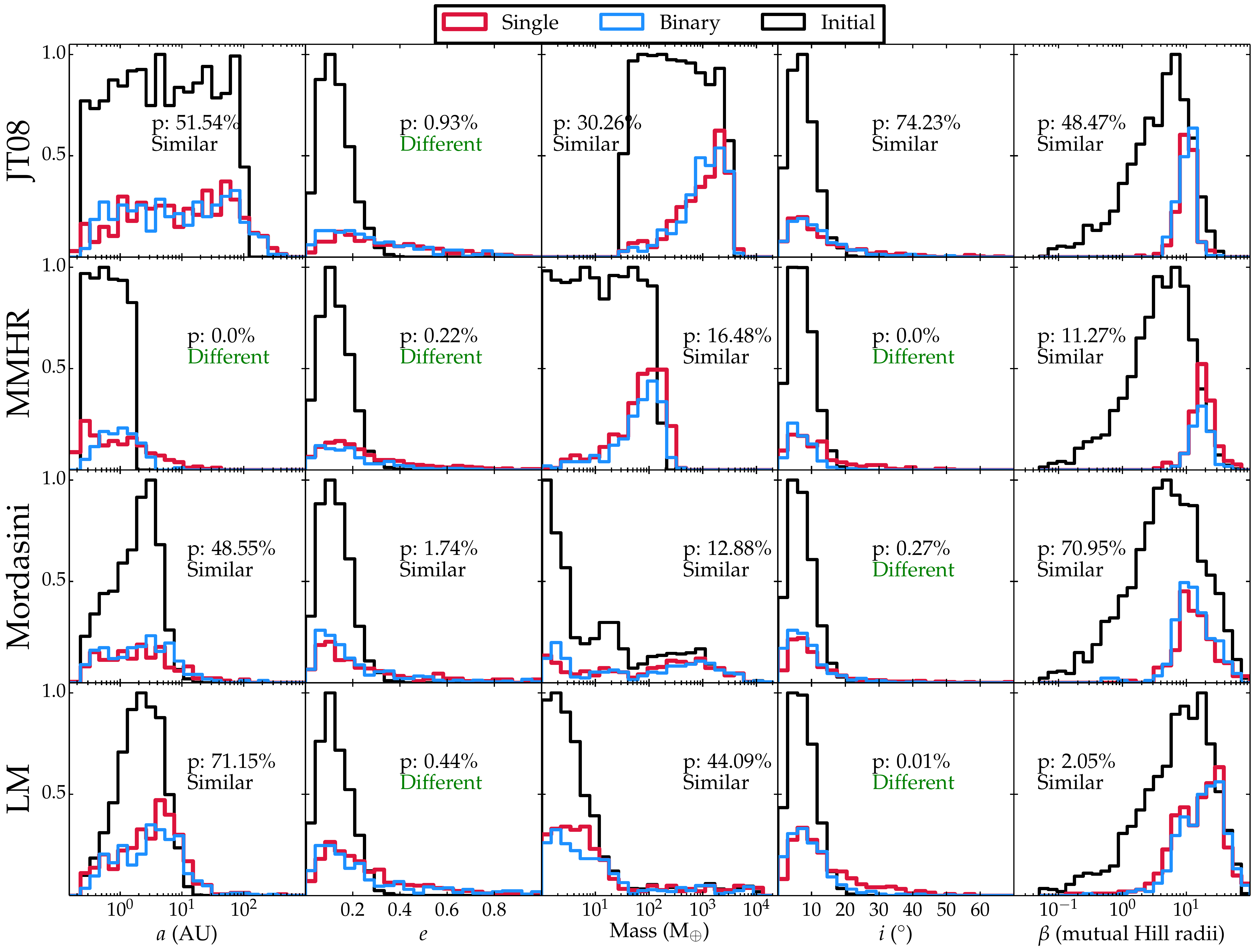}
\caption{ Histograms showing the initial and final distributions of orbital elements for the four planet populations. From top to bottom, the rows depict the JT08, MMHR, Mordasini, and LM populations.  The columns, from left to right, show the $a$, $e$, mass, $i$, and $\beta$ distributions.  The black histogram shows the initial distribution and is created by combining the initial elements for both the binary and single cases.  The single star systems are represented by the the thick red line and circumbinary planets are shown in blue. Final distributions show the populations at 10~Myr.  Each panel is independently normalized by the black initial distribution. We use the Anderson-Darling test to compare the single and binary distributions in each panel. Most of the properties of the planet distributions are minimally affected by the presence of the binary; however, we do see that the CBPs tend to be lower in eccentricity and inclination and that the tighter the initial $a$ distribution, the more different the single and binary populations become (for instance, MMHR).
\label{bvs}}
\end{figure*}

Figure~\ref{bvs}  compares the initial and final distributions of single and circumbinary planetary systems for the four planet populations.  The semi-major axis distributions are broadened as planets are scattered to larger distances  (or smaller, in the single star case, albeit rarely). The peaks of the final eccentricity distributions are similar to the initial distributions, but with a tail at high eccentricities.  Lower mass planets are preferentially lost, leaving dominant populations of higher mass planets.  The inclination distributions (in the binary case, as measured relative to the binary's angular momentum axis) also develop a small tail at higher inclinations, but the majority of planets follow the initial distributions.  Finally, the $\beta$ distributions narrow and shift to higher values, peaking between 10 and 30 \Rhm.  This is similar to the observed packing of \kep\ single star systems reported in \cite{Fang2013} and \cite{Malhotra2015}.

The MMHR planet population has the most variation between the single and binary cases, with the semi-major axis, eccentricity, and inclination all having an Anderson-Darling p-value less than 1\%.  Thus, an initially compact and packed planetary system evolves differently around a binary. Although both binary and single star systems are initialized without very close-in planets, single stars accumulate a sizable population of short period planets.  Independent of central object, the MMHR planets show significant mass accretion due to collisions; the final population has a maximum mass two times higher than the initial maximum mass. JT08 and LM have different eccentricity distributions between single and binary (tending to lower eccentricities in the binary case), and Mordasini and LM have different inclination distributions (tending to lower inclinations in the binary case).  For all populations, the mass and $\beta$ distributions are similar between the single and binary cases. In both cases, the typical separation in $\beta$ is a significantly larger than the minimal value for stability in idealized calculations \citep{Smith2009,Kratter2014}.

\begin{table*}
\centering
\caption{Outcome Fractions: The first column denotes the central object in the system.  The second column shows the integration time in Myr.  The next columns show the percentage of the initial planet populations that remained in the system, were ejected, or suffered planet-star or planet-planet collisions.  The percentages here denote fraction of the total population, which will only be impacted by Poisson noise (32 planets, for our 1000 planet ensemble). } 
\label{tab:pcts}
\begin{tabular}{lccccc}
\hline
System & Int. Time & Stay & Eject & Planet-Star & Planet-Planet\\
 & Myr & \% & \% & \% & \%\\
\hline
JT08\\
\hline
\cite{Juric2008} c10s10 & 100& 26    & 48     & 18    & 8    \\
Single	              & 10   & 28.7  & 51.8   & 14.1 & 5.4   \\
Binary 	              & 10   & 29.3  & 67.3   & 0.8  & 2.6   \\
Binary 	              & 0.5  & 42.3  & 54.3   & 0.8  & 2.6   \\
Binary; $\mu=1$; e=0.1    & 0.5  & 42.0  & 54.0   & 1.5  & 2.5   \\
Binary; $\mu=1$; e=0.5    & 0.5  & 41.6  & 56.5   & 0.7  & 1.2   \\
Binary; $\mu=0.5$; e=0.25 & 0.5  & 46.2  & 50.7   & 0.5  & 2.6   \\
\hline
MMHR\\
\hline
Single	              & 10   & 26.5  & 26.6   & 12.9 & 34.0  \\
Binary 	              & 10   & 19.6  & 64.2   & 0.5  & 15.7  \\
\hline
Mordasini\\
\hline
Single	              & 10   & 28.6  & 48.8   & 14.3 & 8.3  \\
Binary 	              & 10   & 31.0  & 63.3   & 0.5  & 5.2   \\
Binary 	              & 1    & 36.3  & 59.0   & 0.5  & 4.2   \\
Binary; $\mu=1$; e=0.3    & 1    & 31.9  & 63.3   & 1.2  & 3.6   \\
\hline
LM\\
\hline
Single	              & 10   & 48.1  & 30.1   & 9.0  & 12.8  \\
Binary 	              & 10   & 43.5  & 42.7   & 0.4  & 13.4  \\
\hline
\end{tabular}
\end{table*}

\subsection{Differences between planet populations}

We find that planet-star collisions remain roughly constant across all distributions, planet-planet collisions increase with decreased initial semi-major axis range, the number of remaining planets increases with increasing initial mutual Hill radius spacing, and ejections seem to increase with increased average planet mass and planetary compactness, consistent with \cite{Morrison2015}. Despite all systems beginning with 10 planets, there are significant differences in the typical number of planets remaining in a system around the fiducial binary at the end of 10~Myr.  The LM population keeps the most planets (about 4, although half of systems  still have planets on crossing orbits). Mordasini and JT08 each retain about 3, and MMHR systems are left with only 2 on average. In JT08, a massive, widely spaced distribution, the binary plays little role in the rate of planet loss.  The MMHR planets are much more impacted by the binary because of the small semi-major axes: the binary clears out planets very fast. The Mordasini population behaves very similarly to JT08, which is likely due to the presence of a massive planet in most systems (later discussed in Section~\ref{massive}). Finally, the LM planets have a wider initial spacing in mutual Hill radii, so  planet-planet perturbations are weaker, leading to longer instability times.  Additionally, because the planets are mostly low mass, an average planet-planet close encounter will not be able to overcome the system escape velocity.  Thus, to be removed from the system, a planet must interact with a rare giant planet or star, or wait for a relatively rare planet-planet collision.  This leads to the ``long-term'' (several tens of millions of years) survival of planets from this population; note that this is still short compared to the main sequence lifetime of the central stars. There is little difference between the final inclination and eccentricity distributions, although there is a small population of high eccentricity, high inclination planets around binaries for all populations.

\subsection{Impact on stellar binary orbit}

The binary's orbital parameters are impacted very little by the dynamical evolution of the planets; most experienced less than a 1\% change in their orbital characteristics. The most significant changes are seen in the JT08 binary, but even those changes are small. For instance, binary inclinations (measured with respect to the total angular momentum vector of the system) never reach greater than 20\deg\ in the most extreme case, the maximum final stellar eccentricity in initially circular systems is 0.11, and the semi-major axis never changed by more than 0.005~AU (5\%).  

\section{Discussion}\label{disc}

\subsection{Stability of resultant planetary populations}
We have conducted our above analysis after only  10~Myr of evolution. While long term evolution will still occur,  the rate of planet loss (shown in Figure~\ref{nvt}) appears to level off for all populations but LM. A small number of systems in the other populations still have orbit crossings, a sign of ongoing dynamical evolution.  The number of orbit crossings for each of the four distributions in the single and binary cases are shown in Table~\ref{tab:orbitcross}.  We find that, at 10~Myr, the binary systems tend to have fewer orbit crossings as compared to the single star systems.  Additionally, as seen in Figure~\ref{nvt}, the binary systems lose planets faster.  The more rapid onset of planet ejections in the binary case is likely responsible for the reduction in any other kind of collision. Unstable, high eccentricity planets are ejected before they can interact with other planets. All of these effects combined may cause the binary planet populations to be dynamically colder after 10 Myr. If we apply the planet-packing metrics used by \cite{Kratter2014}, we find that our final systems are consistent with being minimally packed, rather than sparse, meaning that the addition of planets in between existing pairs would likely trigger instability. 

Some systems with high numbers of orbit crossings remain stable for nearly the length of the integration, especially for the LM planet population.  In these instances, the kick velocities of the planet-planet encounters are much less than the escape velocity from the system.  These planets will likely remain in the system until a planet-planet collision or a planet-star encounter occurs. When we extend the full complement of single-star LM integrations to $10^8$ years, we find that the rate of planet loss remains constant at about 1 planet per decade of logarithmic time with no sign of reaching a constant number of planets in the system.

We find similar multiplicities to previous \nb\ integrations from \cite{Juric2008}, \cite{Chatterjee2008}, and \cite{Raymond2010}, despite all of these studies being carried out around single stars and with very different initial planet populations and initial multiplicities of 10$+$, 3, and 3, respectively.  All assume  massive planet populations, which is consistent with our finding that initial planet mass has the largest influence on the final system multiplicity, which is discussed in more detail in Section~\ref{massive}.

\begin{table}
\centering
\caption{Orbit Crossings at 10~Myr:  The number of systems with orbit crossings at 10 Myr for both single and binary systems. The first column shows the distribution.  The next three columns show the number of binary systems having no, one, or two or more orbit crossings at 10 Myr. The final three columns show the same information for the single star systems. } 
 \label{tab:orbitcross}
\begin{tabular}{lrrrrrr}
\hline
Dist. & \multicolumn{3}{c}{Binary} & \multicolumn{3}{c}{Single} \\
 & 0 & 1 & 2+ &  0 & 1 & 2+ \\
\hline
JT08      & 82 & 16 & 2  & 73 & 26 & 1  \\
MMHR      & 87 & 12 & 1  & 62 & 28 & 10 \\
Mordasini & 77 & 10 & 13 & 79 & 10 & 11 \\
LM        & 49 &  6 & 45 & 41 & 15 & 44 \\
\hline
\end{tabular}
\end{table}

\subsection{Absence of stellar collisions in circumbinary systems}\label{explanation}

For all circumbinary populations, the dominant form of planet loss is ejection.  Indeed, ejections begin to dominate earlier in the CBP case than in the single star case.  These rapid ejections are triggered when planets cross into the instability region of the binary described in Section~\ref{ics}.  As planets get pumped to higher eccentricities by planet-planet interactions, the pericenter will enter the unstable region around the binary.  Consistent with \cite{Holman1999},  ejections typically occur on timescales of tens to hundreds of planetary orbits after the initial crossing into the instability region.  
For all sets of initial conditions, over 70\% of the planets that had a recorded distance less than 2.23$a_{\textrm b}$ (the empirical \cite{Holman1999} instability boundary) are ejected.  70\% is a rough estimate due to coarse output timesteps in our data that may not record every planet that crossed the instability boundary.  In a high-cadence output test of JT08 CBPs integrated to 100 kyr, more planets are recorded within the instability region, as expected. 

The reduction of planet-star collisions can be understood using intuition gained from the circular restricted three-body problem (CR3BP).  The CR3BP is a well known solution to  the three-body problem in which a test particle orbits in the gravitational potential of two massive bodies on a circular orbit.  While our systems inherently violate the assumptions of the CR3BP due to planets having mass and interacting with one another, we can still gain insight from approximating our systems as multiple instantaneous CR3BPs with each planet as a test particle orbiting the binary, similar to \cite{Moeckel2012} and \cite{Kratter2012}.  One can constrain the allowed orbits of test particles in a binary using the  constant of motion, the Jacobi constant, shown in equation~\ref{eq:cj}.  Here, $n$ is the mean motion of the binary ($n=2\pi/T$ with $T$ being the period, which is unity for our circular binary), $\mu$ is the mass ratio of the stars such that $\mu=1=\mu_1+\mu_2=GM$ and $\mu_1=\mu_2= 0.5$ for our equal mass binary, and $r$ is the position of the planet measured relative to each star. The coordinates and velocities $(x,y,z)$ and $(\dot{x},\dot{y},\dot{z})$ are measured in the inertial frame.  

\begin{equation}\label{eq:cj}
C_J=2n(x \dot{y}-y\dot{x})+2\left( \frac{\mu_1}{r_1}+\frac{\mu_2}{r_2} \right)-\dot{x}^2-\dot{y}^2-\dot{z}^2 
\end{equation}

\begin{figure*}
\centering
\includegraphics[scale=.15]{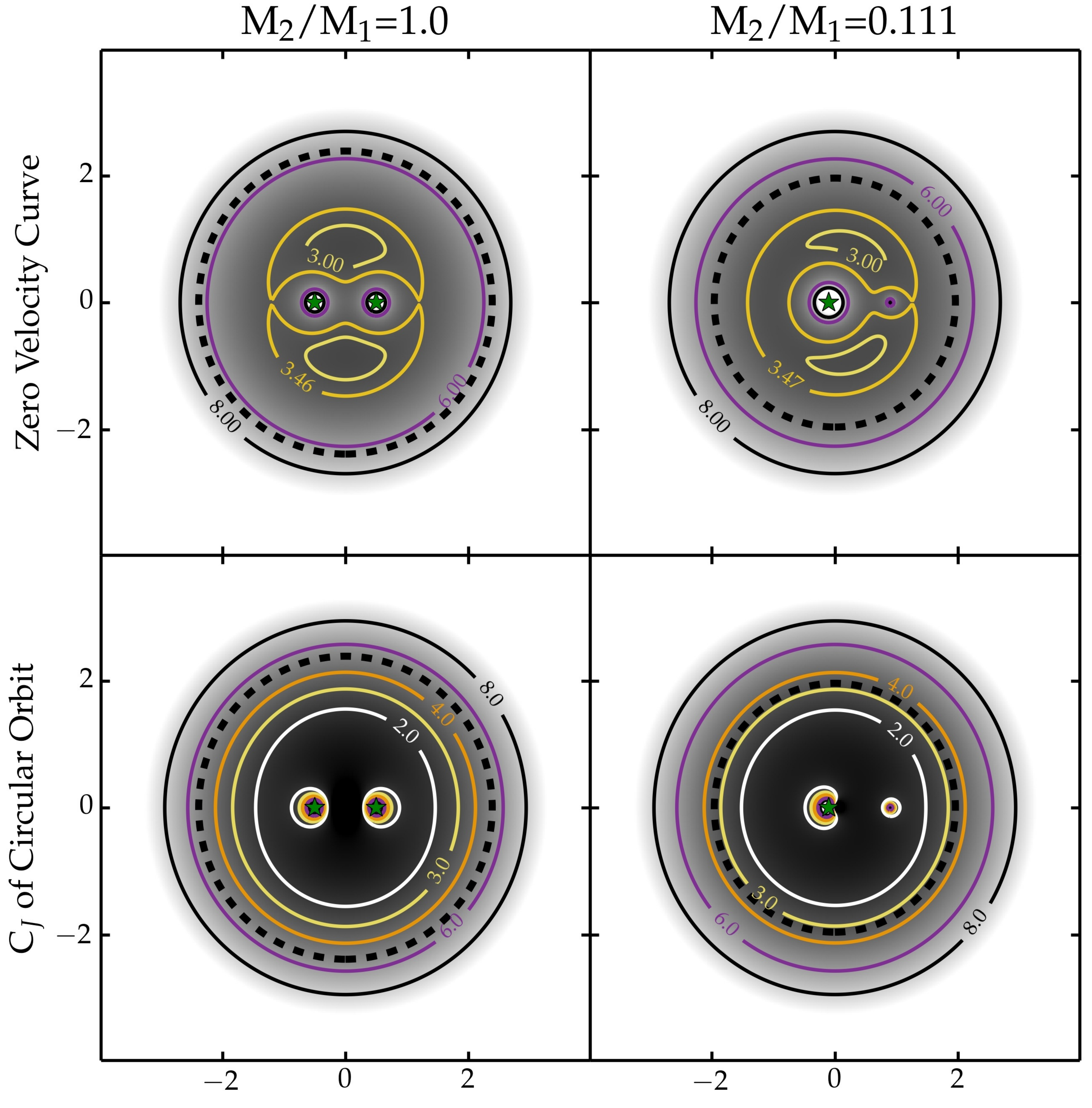}
\caption{ Zero velocity contours (top row) and the Jacobi constants for planets on circular orbits (bottom row) for two mass ratios. Both the gray-scale gradient and colored contours show the value of C$_J$ and are scaled to the same range in all plots. The axes are dimensionless distances scaled to the binary semi-major axis.  The dashed circle in each plot shows the \protect\cite{Holman1999} instability boundary. In the top panel, orbits  cannot cross contours of the same color,  thereby forbidding regions of phase space for a planet.  For C$_J\lesssim3.46$, a planet can collide with the binary; for any higher value of the Jacobi constant, scattering is the only interaction allowed. Comparison with the bottom row reveals that planets beginning on circular orbits exterior to the instability boundary have $C_J$ too large to collide with either star, but may still penetrate the region in which orbits are unstable. We find that while planet-planet perturbations of course violate the CR3BP and change the value of $C_J$ and thus the parameter space available to orbit, interactions in between encounters behave according to the present value of $C_J$. This leads to a much larger number of ejections and a smaller number stellar collisions compared to the single star case.}
\label{zvc}
\end{figure*}

The CR3BP allows us to calculate zero velocity curves for test particle orbits with a given $C_J$, which denote regions in phase space where a given test particle can and cannot orbit.  These are shown in the top row of Figure~\ref{zvc}, where the dashed circle in all panels depicts the \cite{Holman1999} instability boundary for the binary shown.  The bottom row of Figure~\ref{zvc}  shows value of the Jacobi constant for a particle on a circular Keplerian orbit at a particular $(x,y)$ location. A planet with a given Jacobi constant cannot cross a zero velocity contour of the same value.
 
Comparing the plots in the first column of Figure~\ref{zvc}, we see that a planet with $C_J=6$ can reside on a circular orbit outside the instability boundary, but can also just penetrate the unstable region.  If this happens, the planet will be strongly perturbed by the binary and it's orbit will become chaotic. However, a planet with $C_J=6$ cannot collide with either star because the zero velocity contour completely surrounds both stars. Thus a small perturbation from another planet might easily trigger an ejection by sending the planet into unstable region around the binary. However, a small kick would not result in a planet-star collision, because without a substantial change in energy, orbits intersecting the stars are prohibited. Alternatively, a very strong planet-planet encounter causing $C_J$ to decrease to  $\sim3$ would open up the orbital phase space to allow a collision with either star because the zero velocity contours are completely open on both sides.  Because all of the planets in our simulations begin outside the instability boundary, with  $C_J> 4$, they cannot collide with the binary unless an outside perturbation changes the constant of motion. Since most planets have much larger values of $C_J$, very strong planet-planet kicks are necessary to cause collisions. Note that \cite{Szebehely1981} effectively predicted the empirical \cite{Holman1999} boundary based on zero velocity curves. 

In a high-cadence test, most planets that cross into the instability region do not have a $C_J$ such that a stellar collision is possible, yet every planet that does collide has an external interaction that changes the energy in the system such that a collision \emph{is} allowed according to the instantaneous value of $C_J$.  This suggests that despite the inherent simplifications in the above model, $C_J$ provides a useful constraint on available orbital phase space. A large fraction of planetary orbits achieve a $C_J$ that allows them to penetrate the instability region without ever being able to collide with a star. These planets can easily be ejected from the system on short timescales, before they are likely to suffer another planetary encounter that can further decrease $C_J$.  This leads to the $>20$ times decrease in planet-star collisions and can account for some of the increase in ejections.  It is also important to note that a planet with $C_J\le$3.46 is not guaranteed to collide;  this value of the Jacobi constant allows a planet to slip through the zero velocity surface at just the right phase and interact with the stars.  There are still large regions of space that disallow collisions altogether. Direct interactions with the stars are only unconstrained in space for $C_J<$3. 

\begin{figure}
\centering
\includegraphics[scale=.3]{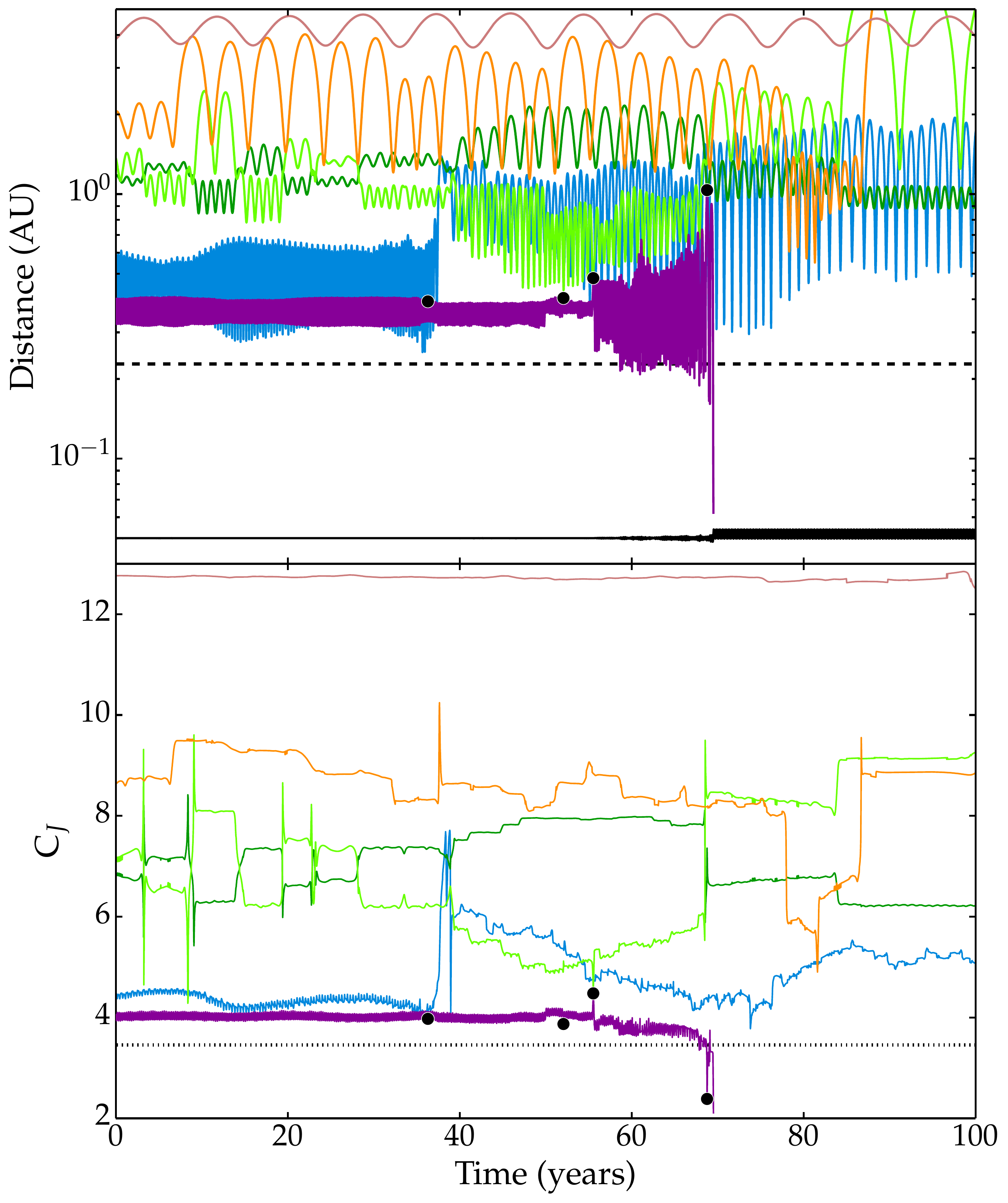}
\caption{ Distance from the barycenter (top) and instantaneous Jacobi constant (bottom) for six planets drawn from the JT08 distribution during the first 100 years of an integration.  The solid black line in the top plot shows the behavior of the secondary star while the dashed line is the \protect\cite{Holman1999} instability boundary. In the bottom plot, the dotted line shows C$_J=3.46$, where the zero velocity curves permit stellar collisions.  The planet denoted by the purple line that begins at 0.33~AU collides with the star at 70 years.  The black points show close encounters between the purple planet and other planets in the system.  After these close encounters, C$_J$ decreases such that the planet can first cross the instability region and then collide with the star. Every planet that collides with the binary has a similar evolution.}
\label{cj}
\end{figure}

Our analysis is consistent with the findings of \cite{Sutherland2016}; they find that ejections are the most common fate of unstable test particles around binaries, and that collisions are primarily with the secondary star (in the second column of Figure~\ref{zvc}, the critical contour by the secondary opens at a higher Jacobi constant than for the primary star). The work from \cite{Sutherland2016} also shows that the trends seen herein hold for moderate eccentricity binaries where the CR3BP is inapplicable. We illustrate how changes in $C_J$ correspond with planetary encounter in Figure~\ref{cj}, where we show the distance from the barycenter and the Jacobi constant for a system drawn from the JT08 distribution.  In this figure, the purple planet collides after undergoing interactions that change the Jacobi constant.

It has been suggested by \cite{Laughlin1997} and \cite{Gonzalez1997} that the collisions of planets with the host stars may provide atmospheric pollution, leading to a measurable metallicity increase. Although the planet-metallicity correlation  is likely dominated by formation effects rather than pollution \citep{Youdin2002}, the effect might still be measurable \citep{Mack2014}. Thus, if circumbinary disks and circumstellar disks have similar planet formation efficiencies, we can speculate that planet-hosting close binaries might show a deficit in pollution signatures as compared to single stars due to the sharp reduction in collisions.

In addition to the reduction of stellar collisions, we see a marked increase in ejections.  Thus, if circumbinary systems form over-packed, we might expect that  a portion of the population of free floating planets, as suggested by \cite{Sumi2011}, originate from binary systems.  \cite{Veras2012} note that the free floating planet population cannot be explained by planet scattering in single star systems alone. The potentially large contribution of free floating planets from binaries is particularly important for microlensing, which is extremely prior-dependent for interpreting detections.   However, known the free-floating planet population is mainly comprised of massive planets, which might be intrinsically rare around close binaries.  Additionally, known CBP hosts, with their short periods, represent a small fraction of the total binary star population \citep{Raghavan2010}.

\subsection{Extent of binary influence}\label{close}

\begin{figure*}
\centering
\includegraphics[scale=0.45]{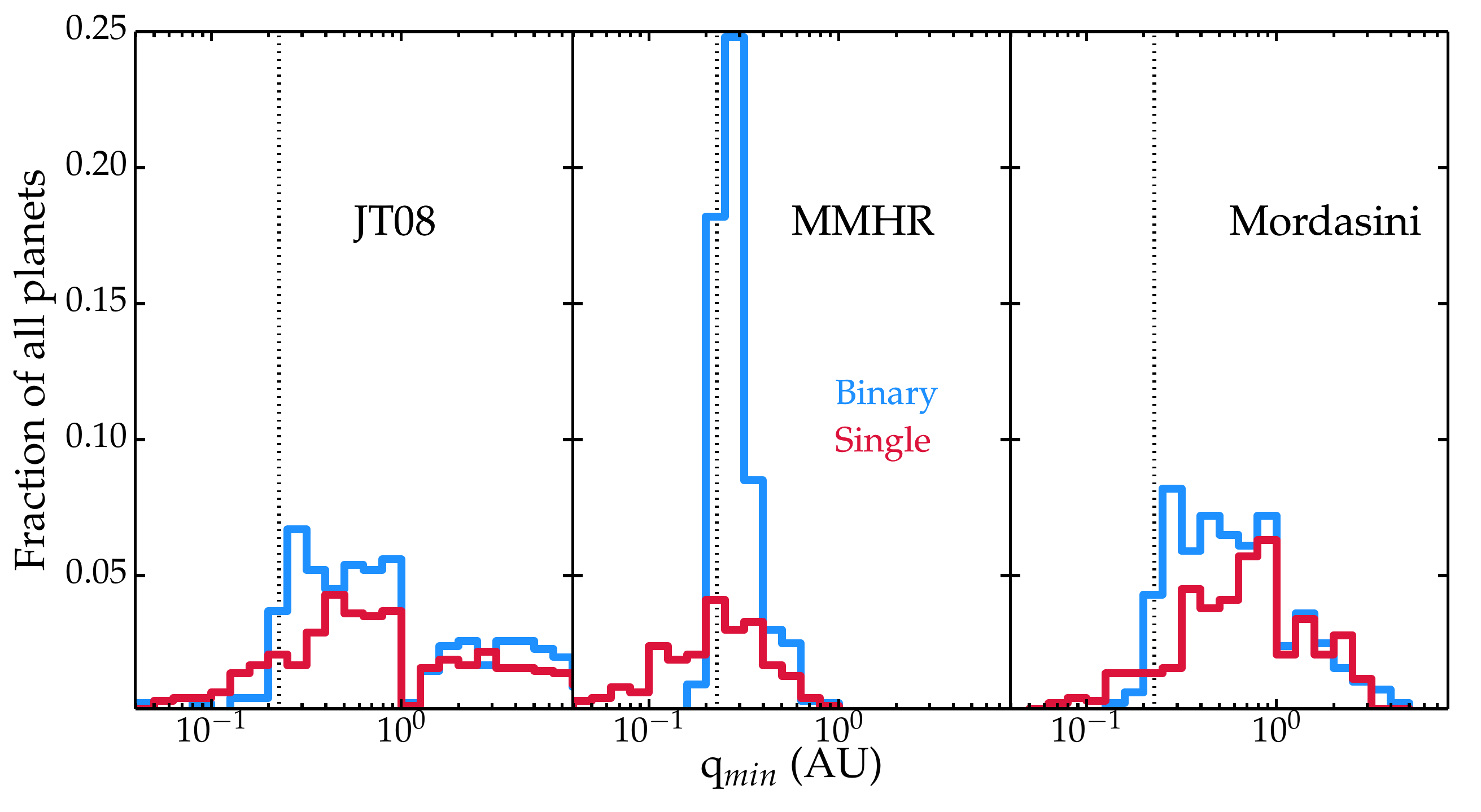}
\caption{ Ejected planets binned by minimum recorded pericenter distance for JT08, MMHR, and Mordasini populations.  The red line shows planets around a single star and the blue line shows planets around a binary.  The vertical dotted line depicts the \protect\cite{Holman1999} instability boundary. The histograms have been normalized to the full 1000 planet sample.  While planets around a single star can get much closer to the star, many more close-in planets are removed from the system due to the binary.  The circumbinary planets dominate ejections in all cases out to about ten times the binary semi-major axis. 
}\label{1au}
\end{figure*}

We highlight the influence of the binary on planet populations in Figure~\ref{1au}, which shows the minimum recorded pericenter distance (relative to the system barycenter) $q$ for ejected planets in the JT08, MMHR, and Mordasini planet populations.  The histograms have been normalized to the total number of planets in the system so that the relative heights of the histograms are indicative of the total population of ejected planets.  For all three populations, planets around a single star have closer pericenter passages, whereas the binary effectively removes planets once they approach the instability region (the dashed line).  The planets in circumbinary systems are preferentially ejected if they pass within $10a_b$, or 1 AU. 

Examining the four planet populations jointly, planets that come within 1~AU of the binary, regardless of initial semi-major axis, have a $\gtrsim$80\% chance of being removed from the system.  Conversely, planets that never come within 1~AU have a 40--80\% chance of \emph{remaining} in the system, depending on the population.   Dynamical evolution leads to 51\% and 76\% of planets crossing within 1~AU despite only 25\% and 29\% of planets initially being at separations closer than 1~AU for the JT08 and Mordasini, respectively.  Thus, we find that the binary has a strong influence on the planet population within order $10a_{\textrm{b}}$. This explains why only the MMHR distribution, which populates this semi-major axis range heavily, shows significantly different final planet statistics between binary and single stars.

Although we have focused on very tight binaries in this study, our results are scalable to wider binaries than we have explored here because planet loss is dominated by ejections. Planetary and stellar radii set some absolute scale, but orbital periods and timescales may be rescaled.  Although planet-planet collision rates decline at larger absolute semi-major axes, these collisions are a small impact on the overall population and would only be smaller when scaled.

\subsection{Role of giant planets in planet multiplicity}\label{massive}
\begin{figure}
\centering
\includegraphics[scale=0.35]{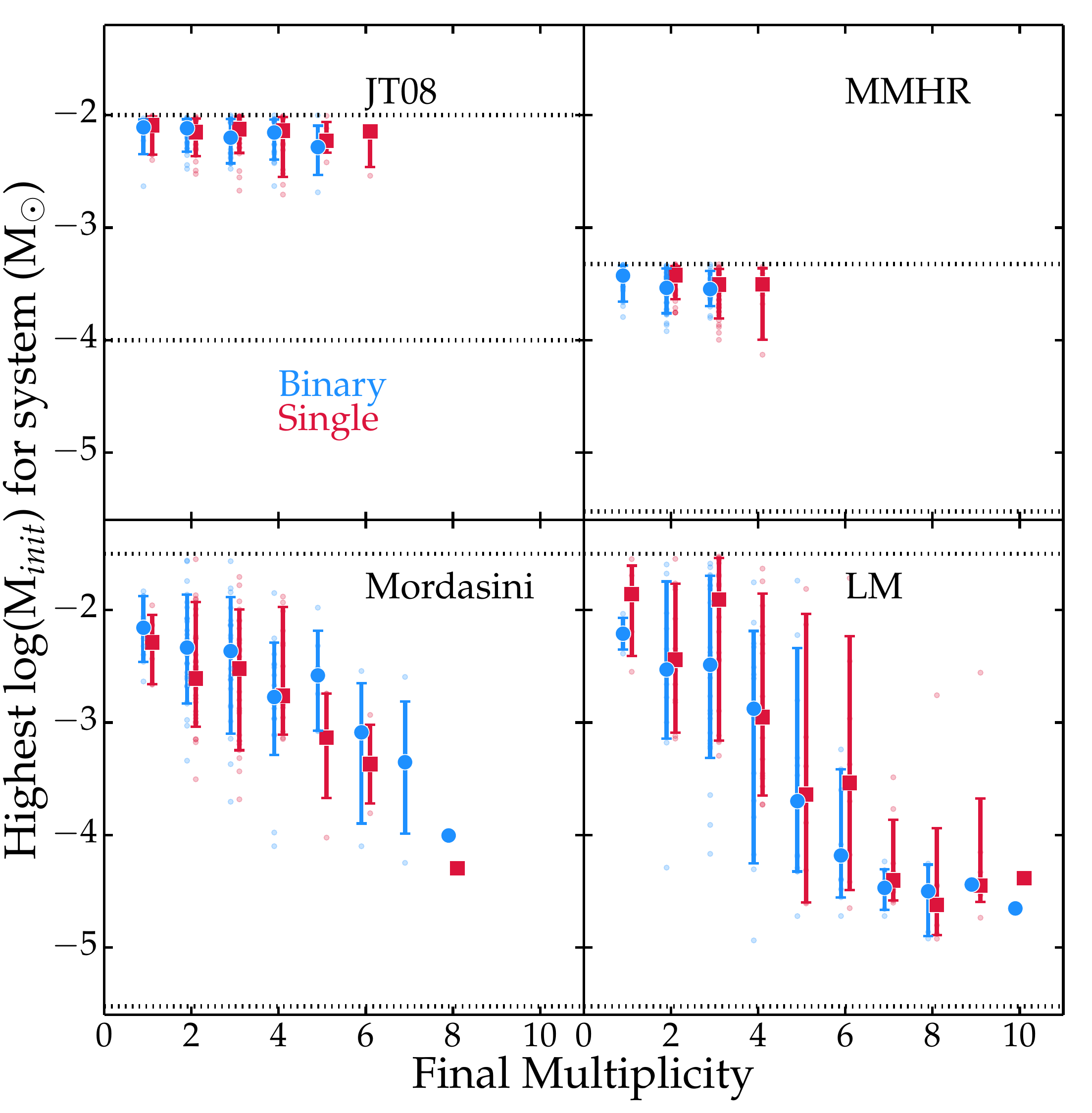}
\caption{ The highest initial mass in the system as a function of multiplicity for the four planet populations. The binary systems are in blue (shifted left) and the single star systems are in red (shifted right).  The small points show the highest mass for an individual system, the symbols (circle for binary and square for single) show the median value of all systems at each multiplicity, and the error bars encompass the 10th-90th percentiles.  The dashed lines show the highest and lowest initial masses for a population.  The colored numbers (blue below and red above) show the number of systems that fall into each multiplicity bin.  We can see that the presence of a Jupiter-mass planet appears to restrict the multiplicity to be less than about 5.  However, the multiplicity of single and binary star systems is overall similar, so intrinsic differences in observed populations are likely due to formation and not scattering.   
\label{nvhm}}
\end{figure}

Observed CBPs lack close-in large planets, as noted in \cite{Martin2015d}.  While this could be a bias of small number statistics, we explore the possibility of it being a dynamical effect. We find that planet multiplicity is a strong function of the highest initial mass in the system, as is shown in Figure~\ref{nvhm}, but there is no statistical difference between planet populations around single stars and binary stars. Specifically looking at our numerical results for Mordasini and LM planetary systems around binaries, systems with Jupiter mass planets undergo very different evolution than their lower mass counterparts.  We find that it is rare for systems beginning with a planet the mass of Jupiter or greater to have a multiplicity greater than five.  Indeed, in the LM case especially, there appears to be a large break in the median highest initial mass in a system between systems containing four and five planets. Systems with four planets have a median highest mass of about 2\mjup, while systems with five planets have a median mass of about 60\mearth, or about  3.5\mnep.  

\begin{figure*}
\centering
\includegraphics[scale=0.3]{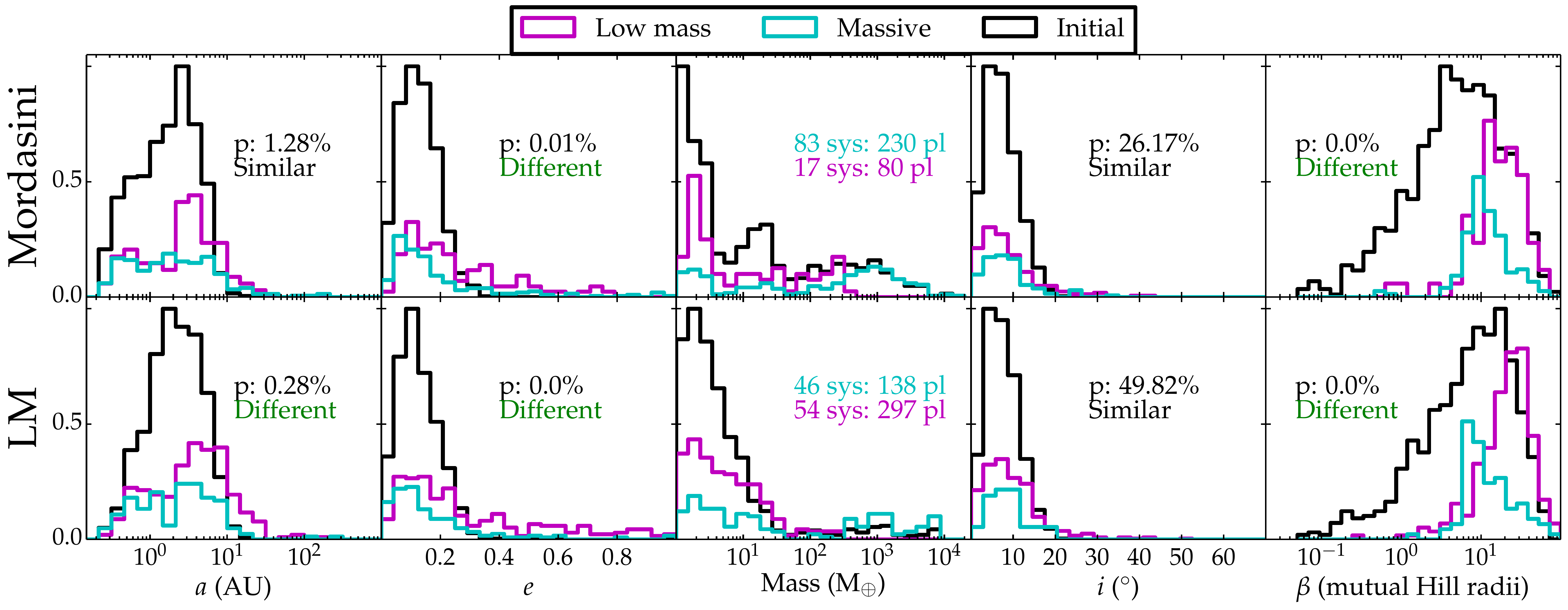}
\caption{ Orbital elements for LM and Mordasini circumbinary populations split by the presence of a massive planet, in the same format as in Figure~\ref{bvs}.  The black histogram shows the initial planet population, the magenta histogram shows the systems containing only planets less than 1~\mjup, and the cyan histogram shows the systems that have a Jupiter-mass planet or larger.  All histograms have been normalized to the maximum of the initial distribution.  The Anderson-Darling p-value comparing the low mass and high mass systems is shown.   There are significant variations in the $\beta$ distributions in particular, showing that the presence of a Jupiter strongly influences the evolution and structure of a planetary system.
\label{splitmass}}
\end{figure*}

 In Figure~\ref{splitmass} we present the resultant orbital elements for the LM and Mordasini CBPs (in the same format as Figure~\ref{bvs}) but split into systems having or lacking a Jupiter.  In the Mordasini case, the resultant eccentricity and $\beta$ distributions are statistically different for the high and low mass cases.  The eccentricity of systems with a Jupiter mass planet is generally lower; both still peak at around 0.1, but the systems without a Jupiter have a small population of high eccentricity planets.  The $\beta$ distributions are also statistically different, with the high mass systems having a peak around 8 and the low mass systems having a peak around 15.  This is consistent with the low mass systems being stable for Gyr timescales, if we apply the \cite{Smith2009} results for single stars.  The LM systems are where the greatest differences are seen.  The eccentricities, semi-major axes, and  $\beta$ distributions are statistically dissimilar.  The high mass systems tend to have smaller semi-major axes, eccentricities, and mutual separations.  The  $\beta$ distribution peaks around 5 for the systems with massive planets and 30 for systems without. These characteristics all suggest that the absence of a high mass population of circumbinary planets is not the result of different dynamical evolution.  The increase in dynamical evolution in the presence of a giant planet was also seen by \cite{Hands2015}.

Systems lacking giant  planets are more widely spaced in mutual Hill radii. Because the outcome of planet-planet perturbations is controlled by the more massive body,  systems without a massive planet have a larger probability of non-catastrophic planet-planet interactions.  Thus, the systems can stay intact at high multiplicity.  However, as will be discussed in Section~\ref{tran}, we find no correlation between intrinsic multiplicity and the number of transits seen in a system for the populations studied here.  Therefore, the lack of observed giant planets stems from either small number statistics or formation.

\subsection{Observables}\label{tran}
In order to make a rough comparison with \kep\ detections, we make a simplified calculation of the number of planets that would transit based on the following limiting assumptions. We assume that the systems are seen along $i=90$ (edge on) for the binary systems. We also neglect planets' orbital evolution over our assumed 5 year ``mission lifetime'' and limit ourselves to planets having at most a 2 year period. \cite{Martin2015b} show that, given enough time, nearly all circumbinary planets are expected to transit due to precession effects, so these results are only valid for short duration monitoring. We find that 10--30\% of systems have at least one transiting planet; only 1--5\% show more than one transit.  The number of systems showing a given number of transits is shown in Table~\ref{tab:transit}. There is no correlation between observed multiplicity and intrinsic multiplicity. To provide a comparable sample around the single stars, we assume an equatorial line of sight and calculate transits.  While this is an oversimplification, as stars are randomly oriented with respect to the observer, randomly choosing lines of sight would only decrease the number of observed transits and we want to compare transit rates for comparably-aligned systems. We find a slightly higher number of transits for the single star systems, but the planets around a single star can reside closer to the central star in a stable orbit and are therefore more likely to transit. Similarly, the number of transits is a function of the compactness of a planet population's semi-major axis distribution. We find no correlations with the probability of a transit and planet mass, but the mass of transiting planets is roughly consistent with the mass of the initial distribution. This confirms that the dearth of giant planets on close-in orbits is not the result of different scattering behavior around binaries.  The fraction of observed single planet systems to observed multi-planet systems is consistent with the known \kep\ systems from \cite{Batalha2013}. 

\emph{Kepler-47} We also investigate the ability of our simulations to create a system like Kepler-47, which has three nearly unstable planets close to the central binary (binary period of 7.5 days and planet periods 49.5, 187, and 303 days \citep{Welsh2015}).   While not common, we find that both the LM and Mordasini populations finish the 10Myr simulation with a handful of moderate multiplicity, tightly packed (both dynamically and physically) systems. However, the majority of three planet systems have average semi-major axes much larger than the true Kepler-47.  Thus, Kepler-47 could be the remnant of a system sculpted by dynamical evolution but would require rather extreme initial conditions.

\begin{table*}
\centering
\caption{Number of transits: The number of systems showing transits.  The first column shows the planet population.  Each subsequent pair of of columns (binary on the left and single star on the right) shows the number of systems with a given number of transits. }
 \label{tab:transit}
\begin{tabular}{lcccccccc}
\hline
{Dist.} & \multicolumn{2}{c}{0} & \multicolumn{2}{c}{1} & \multicolumn{2}{c}{2} & \multicolumn{2}{c}{3+} \\
 & Binary & Single  & Binary &  Single  & Binary &  Single  \\
\hline
JT08 & 88 & 87 & 12 & 13 & 0 & 0 & 0 & 0 \\
MMHR & 81 & 66 & 15 & 30 & 4 & 4 & 0 & 0 \\
Mordasini & 75 & 71 & 24 & 28 & 1 & 1 & 0 & 0 \\
LM & 77 & 80 & 21 & 12 & 2 & 7 & 0 & 1 \\
\hline
\end{tabular}
\end{table*}

\section{Conclusions}

We have performed N-body simulations of planet-planet scattering around single and binary stars to tease out the influence of a central binary on the dynamical evolution of the system. Our modified version of the \merc\ code has been released online.  Our most important findings are as follows:

\begin{enumerate}
\item The average loss rate for planets is very similar between single star planetary systems and CBPs for a range of initial orbital distributions, though there is a weak dependence on the compactness of the initial semi-major axis distribution.  Planets packed closer to the binary will be more perturbed.
\item The loss method between single star and binary systems is very different.  Circumbinary systems always have far more ejections than the single star planetary systems, and both planet-planet and planet-star collisions are suppressed around binaries (planet-star collisions often by an order of magnitude).  Using intuition based off the CR3BP, these reductions are expected because the orbital phase space in which planets are perturbed and rapidly ejected is much larger than the phase space allowing stellar collisions.  We speculate that the reduction of collisions in circumbinary systems may lead to a measurably lower atmospheric metallicity in close binaries than in single stars or wide binaries.
\item There are few differences in the final orbital distributions of planets around single and binary stars.  The final planet populations have characteristics that depend mostly on the initial populations, not on the central object.  We see no evidence for a planet pileup around the binary instability boundary. We also find that systems similar to Kepler-47, while not common, are not prohibited by scattering.
\item Systems with a giant planet evolve differently than those without one. The highest multiplicity systems do not have massive planets.  However, the presence of a giant planet has a similar impact on single and binary star systems.
\end{enumerate}

We have shown that intrinsic differences in the populations of CBPs and exoplanets around single stars likely arise from differences in formation or disk-driven orbital evolution. We see no evidence that the lack of observed giant planets nor the pile-up of planets around the binary instability boundary can be attributed to planet-planet and planet-star scattering. And yet, the binary does impact planets that come within roughly a factor of 10 of the binary semi-major axis.  Planets born in-situ on close-in orbits are most likely to evolve differently around binary and single stars. However, this parameter space is where planet formation is most likely inhibited around binaries due to the excitation of the disk and planetesimal eccentricities close to the stars.  If planet formation around binaries is very efficient, circumbinary systems might be responsible for a population of free floating planets. Thus, while dynamical evolution may not hold the key to creating intrinsic differences in circumbinary and single star planetary systems, it may provide the crucial observational evidence we need to understand these differences.

\section*{Acknowledgements}
 We sincerely appreciate the comments of our reviewer, Alexander James Mustill, and his work to improve this paper. Our gratitude to Dimitri Veras and Renu Malhotra for their insightful and encouraging comments on an early draft. RAS is supported by the National Science Foundation under Grant No. AST-1410174 and Grant No. DGE-1143953. KMK is supported by the National Science Foundation under Grant No. AST-1410174.  AS is supported by the European Union through ERC Grant No. 279973.  The numerical simulations presented herein were run on the El Gato supercomputer, which is supported by the National Science Foundation under Grant No. 1228509.






\bsp	
\label{lastpage}
\end{document}